\documentclass[aps,prb,superscriptaddress,floatfix,twocolumn,longbibliography]{revtex4-2}

\usepackage{amsmath,dcolumn,bm,amsthm,tabularx,subfigure}
\usepackage[colorlinks=true,urlcolor=blue,citecolor=blue,linkcolor=blue]{hyperref}
\usepackage{mathrsfs,times}
\usepackage{bbold,dsfont}
\usepackage{graphicx,graphics,color}
\usepackage{mathtools,nicefrac}
\usepackage{slashed}
\usepackage{amsfonts,amssymb}
\usepackage{multirow}
\usepackage{booktabs}
\usepackage{mathptmx}
\allowdisplaybreaks

\newcommand{\beq}{\begin{equation}}
\newcommand{\eeq}{\end{equation}}
\newcommand{\bea}{\begin{eqnarray}}
\newcommand{\eea}{\end{eqnarray}}

\begin{document}

\title{Universal Entanglement Pattern Formation via a Quantum Quench}

\author{Lihui~Pan}
\affiliation{Key Laboratory of Artificial Structures and Quantum Control (Ministry of Education), School of Physics and Astronomy, Shanghai Jiao Tong University, Shanghai 200240, China}

\author{Jie Chen}
\email{chenjie666@xhu.edu.cn}
\affiliation{School of Science, Key Laboratory of High Performance Scientific Computation, Xihua University, Chengdu 610039, China}

\author{Chun~Chen}
\email{chunchen@sjtu.edu.cn}
\affiliation{Key Laboratory of Artificial Structures and Quantum Control (Ministry of Education), School of Physics and Astronomy, Shanghai Jiao Tong University, Shanghai 200240, China}

\author{Xiaoqun~Wang}
\email{xiaoqunwang@zju.edu.cn}
\affiliation{School of Physics, Zhejiang University, Hangzhou 310058, Zhejiang, China}
\affiliation{Collaborative Innovation Center of Advanced Microstructures, Nanjing University, Nanjing 210093, China}

\date{\today}

\begin{abstract}
We identify a universal short-time structure in symmetry-resolved entanglement dynamics---the entanglement channel wave (ECW)---arising from the decomposition of entanglement into conserved-quantum-number sectors that host robust, channel-specific patterns. Focusing on domain-wall melting, we conduct a systematic investigation across three paradigmatic classes of many-body systems: U(1) fermions, U(1) bosons, and SU(2) spinful fermions. For each class, we explore four distinct regimes defined by the presence or absence of interactions and disorder, employing both the Krylov-subspace iterative method and the correlation matrix approach. The ECW emerges universally across all cases, establishing its independence from particle statistics, interaction strength and disorder. In free fermions, the ECW formalism further enables analytical determination of the correlation matrix spectrum. The subsequent melting of the ECW exhibits symmetry- and statistics-dependent signatures, revealing finer structures in the growth of symmetry-resolved entanglement.
\end{abstract}

\maketitle

\section{Introduction}

Noether’s theorem establishes that every symmetry is associated with a conserved quantity, and such conserved quantities naturally induce a decomposition of the Hilbert space. This, in turn, allows the entanglement entropy to be resolved into contributions from sectors labeled by the corresponding conserved charges—a framework known as symmetry-resolved entanglement entropy (SREE) \cite{goldstein2018,feldman2019b,xavier2018,capizzi2021,calabrese2021,estienne2021,kusuki2023a,fossati2023,turkeshi2020,parez2021,parez2021a,bonsignori2019a,azses2020a,fraenkel2021a}. As measurements of entropy become increasingly feasible in cold-atom platforms \cite{brydges2019,daley2012,lukin2019,islam2015,elben2020}, SREE is expected to draw growing experimental interest \cite{neven2021,vitale2022}.

Using conformal field theory (CFT), studies of critical ground states have uncovered a striking universal feature—entanglement equipartition—where the entanglement within each symmetry sector is independent of the magnitude of the associated conserved charge. This property has been demonstrated in systems with U(1) symmetry \cite{xavier2018}, non-Abelian symmetries \cite{calabrese2021}, disorder \cite{turkeshi2020}, and even non-Hermitian settings \cite{fossati2023}. CFT methods can also describe certain excited states \cite{capizzi2020a}, where the leading-order behavior retains the equipartition structure while subleading corrections acquire a dependence on the conserved charge $q$.

Beyond CFT, quench dynamics of free fermions have been analyzed using alternative approaches \cite{parez2021,parez2021a}. For specific initial states—such as the Néel and dimer states—when the charge lies close to its average value, an effective entanglement equipartition survives, but with a non-universal correction proportional to $(\Delta q)^2$ that is absent in CFT predictions. Another beyond-CFT feature is a delay in the onset of SREE growth, scaling linearly with $|\Delta q|$.

Then, is there a universal symmetry-resolved entanglement pattern that can be obtained without relying on CFT? In this work, we employ SREE in place of the total entanglement entropy \cite{berkovits2012,pouranvari2015} or the number entropy \cite{ghosh2022a,kiefer-emmanouilidis2020} to study domain wall melting \cite{capizzi2023a,hauschild2016}. We find that, although disorder and interactions can significantly affect the SREE at long times, a universal short-time pattern emerges across different systems—independent of CFT descriptions—which we refer to as the entanglement channel wave (ECW) \cite{chen2023c}.

In Sec.~\ref{sec2}, we present and numerically verify the ECW picture in fermionic systems, and further employ it to derive the short-time behavior of the spectrum of the correlation matrix for free fermions. In Sec.~\ref{sec3}, we extend the ECW picture to bosonic systems and SU(2) SREE, discussing both the disappearance of ECW over time and the long-time behavior of SREE in different systems. Finally, in Sec.~\ref{sec4}, we provide our conclusions and outlook.

\section{Emergence of entanglement entropy patterns: The entanglement channel wave phenomenon} \label{sec2}

In this section, we explore the short-time quench dynamics of symmetry-resolved entanglement entropy in a one-dimensional system of spinless fermions. The system is governed by the disordered spinless fermionic Hubbard (dFH) model \cite{hauschild2016} on a lattice of length $L$ (with $L$ an even integer) and periodic boundary conditions (PBCs): 
\begin{align}
\hat{H}_{\text{dFH}} &= J\sum_{i} ( c_i^\dagger c_{i+1} + \text{H.c.})+ \sum_{i}\mu_i (2n_i - 1) \nonumber\\
&+ \sum_{i} \frac{U}{2} (2n_i - 1)(2n_{i+1} - 1), \label{dFH} 
\end{align}
where $c_{i}$ and $c_{i}^{\dagger}$ are fermionic creation and annihilation operators satisfying anticommutation relations for spinless fermions. The number operator $n_{i}= c_{i}^{\dagger}c_{i}$ measures the local occupation on site $i$, with site indices defined modulo $L$ due to PBCs. The first term describes nearest-neighbor hopping with amplitude $ J $. The second term encodes a site-dependent disordered chemical potential $\mu_{i}\in[-\mu,\mu]$, which tunes the local particle density. The third term corresponds to a nearest-neighbor density-density interaction with strength $U$. Via the Jordan-Wigner transformation \cite{mbeng2024}, this model can be mapped onto the $XXZ$ spin chain with a random external field \cite{znidaric2008}. The system preserves a global U(1) symmetry associated with the conservation of total particle number $\hat{Q}=\sum_{i}n_{i}$.

We investigate the dynamics of symmetry-resolved entanglement entropy for initial pure states $\rho=\left|\psi(0)\right\rangle\left\langle\psi(0)\right|$ \cite{goldstein2018,capizzi2021}. The symmetry condition $\left[\rho(t),\hat{Q}\right]=0$ implies a block decomposition of the Hilbert space as $\mathcal{H}^{q}=\bigoplus_{q_{A}} \mathcal{H}_{A}^{q_{A}}\bigotimes\mathcal{H}_{B}^{q-q_{A}}$ arising from the charge partitioning $\hat{Q}=\hat{Q}_{A}+\hat{Q}_{B}$. Accordingly, the reduced density matrix of subsystem $A$ takes the form $\rho_{A}(t)=\bigoplus_{q_{A}}\rho_{A,q_{A}}(t)$, where each block $\rho_{A,q_{A}}(t)$ defines a normalized sector density matrix $\tilde{\rho}_{A,q_{A}}(t)=\rho_{A,q_{A}}(t)/\mathrm{Tr}(\rho_{A,q_{A}}(t))$. Following Refs.~\cite{goldstein2018,capizzi2021}, the symmetry-resolved Rényi entropy is defined as: $S^{q_{A}}_{\alpha}(t)=\ln\mathrm{Tr}(\tilde{\rho}^{\alpha}_{A,q_{A}}(t))/(1-\alpha) $, and the corresponding von Neumann entropy as the $\alpha\rightarrow1$ limit, $S^{q_{A}}_{\rm vN}(t)=S^{q_{A}}_{1}(t)=-\mathrm{Tr}\tilde{\rho}_{A,q_{A}}(t)\ln(\tilde{\rho}_{A,q_{A}}(t)) $. We consider an inhomogeneous domain-wall initial state, $\left|\bullet^{\bigotimes L/2}\circ^{\bigotimes L/2}\right\rangle$. Subsystems $A\equiv\{\text{initially filled sites}\}$ and $B=\bar{A}$ the remaining empty ones. Under U(1)-symmetric dynamics governed by Eq.~\eqref{dFH}, we track the time evolution of the symmetry-resolved von Neumann entropy for fermions. Using the Krylov-subspace iterative method \cite{paeckel2019}, we uncover a novel dynamical structure in the symmetry-resolved entanglement profile, which we refer to as the \lq\lq entanglement channel wave.''

\subsection{Physical picture of the entanglement channel wave}

\begin{figure*}[!t]
\includegraphics[width=\textwidth, height=\textheight, keepaspectratio]{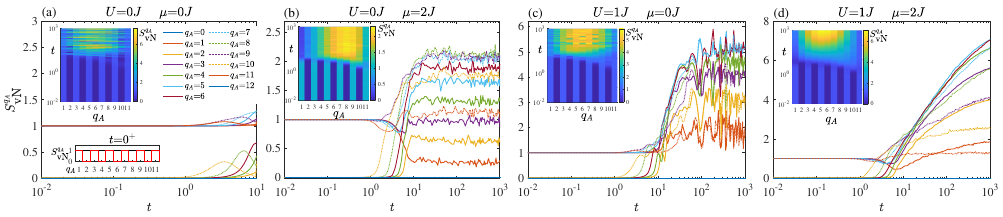}
\caption{(a)-(d) Time evolution of the SREE \( S_{\text{vN}}^{q_A} \) (in units of \( \log 2 \)) for half-chain entanglement in a system of spinless fermions with \( L = 24 \), initialized in a domain-wall state. The entropies are computed for charge sectors \( q_A = 0 \) to \( 12 \). Each panel corresponds to a different combination of interaction strength \( U \) and disorder strength \( \mu \). The time axis is shown on a logarithmic scale. Top-left insets display the emergence and persistence of a parity-dependent entropy pattern (\( S_{\text{vN}}^{q_A} = 0 \) for even \( q_A \), \( \log 2 \) for odd \( q_A \)) over the time range \( Jt \in [10^{-2}, 10^{0}] \). Bottom-left inset of (a) illustrates the ECW pattern at \( t = 0^+ \), consistent with the theoretical prediction in Eq.~(\ref{entropy}). 
For disordered cases, results are averaged over 100 disorder realizations. \label{fermi}} 
\end{figure*}

We compute the symmetry-resolved von Neumann entropy following a quantum quench from a domain-wall initial state, examining four representative cases: (i) free and clean $(U=0,\mu=0)$, (ii) free with disorder $(U=0,\mu\neq0)$, (iii) interacting and clean $(U\neq0,\mu=0)$, and (iv) interacting with disorder $(U\neq0,\mu\neq0)$. In the disordered scenarios, the entropy in each charge sector is averaged over 100 disorder realizations. The resulting time evolution of the symmetry-resolved entanglement is shown in Fig.~\ref{fermi}.

\begin{figure*}[!t]
\includegraphics[width=0.8\textwidth, keepaspectratio]{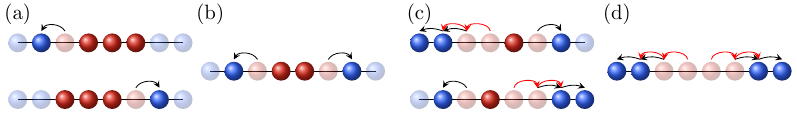}
\caption{(a)-(d) Shortest-distance configurations contributing to entanglement in the fermionic system with \( L = 8 \), shown for charge sectors \( q_A = 3, 2, 1, 0 \), respectively. Red (blue) dots denote particles in subsystem \( A \) \( (B) \). PBCs are imposed, and subsystem \( A \) has been shifted to the center for clarity. In each configuration, particles that have escaped subsystem \( A \) preferentially occupy the edge of subsystem \( B \). \label{L8_Fermi}}
\end{figure*}

At $t=0$, the SREE of subsystem $A$ vanishes across all charge sectors. This is because the domain-wall initial state yields a diagonal reduced density matrix $\rho_{A}(0)$ with a single nonzero entry. However, in an infinitesimally short time $(Jt=0^+)$, a clear parity-dependent structure emerges: the SREE takes the value $0$ $(\log 2)$ for even (odd) channel, as shown in the top-left insets of each panel in Fig.~\ref{fermi}. More generally, as we demonstrate later, the symmetry-resolved Rényi entropy at $t\rightarrow0^{+}$ $q_{A}\neq 0 $ for any nonzero charge sector $q_{A}$ and any Rényi index obeys a universal parity quantization: 
\bea
 S^{q_{A}}_{\alpha}(0^{+}) =\frac{1-(-1)^{q_{B}}}{2}\log 2, \label{entropy}
\eea
where $q_{B}=L/2-q_{A}$ denotes the conserved charge in subsystem $B$ under half-chain bipartitioning as shown in the bottom inset of Fig.~\ref{fermi}(a). This nonlocal, parity-dependent entanglement imbalance---analogous in spirit to the alternating local density in a $Q=\pi$ charge-density-wave (CDW) order, is characterized by a local density modulation between even and odd sites---defines what we call the entanglement channel wave \cite{chen2023c}.  

To understand the mechanism behind the ECW phenomenon, we begin by analyzing the free and clean system. In this setting, the reflection symmetry $\mathcal{R} = \mathcal{R}_{A} \bigotimes \mathcal{R}_{B}$ (where $\mathcal{R}_{\nu}$ reverses the configuration in subsystem $\nu$) is preserved and imposes a constraint on the reduced density matrix:
\begin{align}
(\rho_{A,q_A})_{\alpha,\alpha'} &=\sum_{\beta\in\mathcal{H}_{B}^{q_{B}}}\left\langle\alpha\beta \right|\left.\psi(t)\right\rangle \left\langle \psi(t)\right|\left. \alpha'\beta\right\rangle \nonumber      \\
&=\sum_{\beta\in\mathcal{H}_{B}^{q_{B}}}\left\langle\mathcal{R}_{A} \alpha\mathcal{R}_{B}\beta \right|\left.\mathcal{R}\psi(t)\right\rangle\nonumber\\
&\times
\left\langle \mathcal{R}\psi(t)\right|\left. \mathcal{R}_{A}\alpha'\mathcal{R}_{B}\beta\right\rangle \nonumber      \\
&=\sum_{\beta\in\mathcal{H}_{B}^{q_{B}}}\left\langle\mathcal{R}_{A} \alpha\beta \right|\left.\psi(t)\right\rangle\left\langle \psi(t)\right|\left. \mathcal{R}_{A}\alpha'\beta\right\rangle\nonumber\\
&= (\rho_{A,q_A})_{\overline{\mathcal{R}_A\alpha},\overline{\mathcal{R}_A\alpha'}} \prod_{\mu=\alpha,\alpha'} \epsilon_{\mathcal{R}_A\mu}^{\overline{\mathcal{R}_A\mu}},\label{reflection}
\end{align}
where the phase factors $\epsilon_{\mathcal{R}_{A}\mu}^{\overline{\mathcal{R}_{A}\mu}}=(-1)^{N_{\text{swap}}(\mathcal{R}_{A}\mu\rightarrow\overline{\mathcal{R}_{A}\mu})}$ arise from reordering fermionic basis states, with $N_{\text{swap}}$ counting the number of sign changes due to fermionic exchange. Eq.~\eqref{reflection} reveals a symmetry-enforced pairing structure within the reduced density matrix, where each matrix element $\rho_{\alpha,\alpha'}$ is equal (up to a sign) to its reflected counterpart $\rho_{\overline{\mathcal{R}_A \alpha}, \overline{\mathcal{R}_A \alpha'}}$. This structure holds universally, except for $\mathcal{R}_A$-invariant states (satisfying $\overline{\mathcal{R}_A \alpha} = \alpha$), whose behavior may be affected by the specific spatial arrangement of particles. The short-time behavior of the symmetry-resolved entanglement entropy for charge $q_A$ is governed by the leading-order contributions in $t$ to the reduced density matrix $\rho_{A,q_A}(t)$. Specifically, the dominant terms in the matrix elements $\left\langle \alpha\right|\rho_{A,q_{A}}(t)\left|\alpha'\right\rangle=\sum_{\beta\in \mathcal{H}^{q_{B}}_{B}}\left\langle \alpha\beta\right|\left.\psi(t)\right\rangle\left\langle\psi(t)\right. \left|\alpha'\beta\right\rangle$ originate from those $\left\langle \alpha\beta \middle| \psi(t) \right\rangle$ with the lowest order in $t$. By Taylor expanding the evolution operator $\exp(-i\hat{H}t)$, one finds that the leading-order term is controlled by the \textit{distance} between the initial state $\left|\psi(0)\right\rangle$ and the final configuration $\left|\alpha \beta\right\rangle$, where distance is defined as the minimal number of hopping steps required to connect the two configurations.
Configurations in $\mathcal{H}_{A}^{q_A} \bigotimes \mathcal{H}_{B}^{q_B}$ that minimize this distance yield the leading-order terms in $\rho_{A,q_A}(t)$. These configurations \( |\alpha^s \beta^s\rangle \), corresponding to particles tunneling from the edge of subsystem \( A \) into the edge of subsystem \( B \), are depicted in Fig.~\ref{L8_Fermi}. For $q_A \neq 0$, these shortest-distance configurations exhibit two key properties:

\begin{enumerate}
    \item Multiplicity: The number of such configurations within the charge channel \( q_A \) is given by \( 2^{(1-(-1)^{q_B} )/2} \).
    \item Pairing: The sets \( \{\alpha^s\} \in \mathcal{H}_A^{q_A}\) and \(\{\beta^s\} \in \mathcal{H}_B^{q_B} \) form one-to-one pairs within these configurations.
\end{enumerate}

The first property directly follows from the reflection symmetry constraint in Eq.~\eqref{reflection}. When \( q_B \) is even, the subsystem configuration \( |\alpha\rangle \) tends to be self-dual under reflection due to edge-tunneling symmetry, resulting in a single valid configuration. When \( q_B \) is odd, the reflection symmetry leads to two distinct configurations.

The second property implies that off-diagonal elements of the reduced density matrix \( \rho_{A,q_A}(t) \) are of higher order in $t$ than the diagonal ones, and can thus be neglected in the limit \( t \to 0^+ \). Therefore, the only nonzero elements of the normalized reduced density matrix $\tilde{\rho}_{A,q_A}(0^+)$ are:
\begin{itemize}
    \item For even \( q_B \), \( \tilde{\rho}_{A,q_A}(0^+) \) possesses a single diagonal element equal to 1.
    \item For odd \( q_B \), \( \tilde{\rho}_{A,q_A}(0^+) \) has two diagonal elements, each equal to \( 1/2 \).
\end{itemize}

As a result, the SREE exhibits a parity-dependent structure. For $q_A \ne 0$, it takes the form of Eq.~\eqref{entropy}, yielding $\log 2$ for odd $q_B$ (two equally probable states) and zero for even $q_B$ (single configuration). For $q_A = 0$, the entropy vanishes due to the triviality of the corresponding Hilbert space $\mathcal{H}_A^{q_A = 0}$ (and similarly for $\mathcal{H}_B^{q_B = 0}$), which implies that $S^{q_A=L/2}_\alpha(t)$ is also zero, since $\text{Tr}(\tilde{\rho}_{A,q_A}^\alpha(t)) = \text{Tr}(\tilde{\rho}_{B,L/2 - q_A}^\alpha(t))$.

Finally, since the interaction and potential terms in the Hamiltonian are diagonal in the occupation-number basis, they do not induce particle hopping and thus contribute only higher-order corrections in $t$. As our analysis focuses on the leading-order terms of $\rho_{A,q_A}(t)$, these diagonal terms do not affect our conclusions. We therefore find that the parity quantization of the SREE---expressed in Eq.~\eqref{entropy}---remains robust even in the presence of interactions and disorder.

\subsection{Correlation properties of free fermions induced by the ECW}

\begin{figure*}[!t]
\includegraphics[width=0.4\textwidth, keepaspectratio]{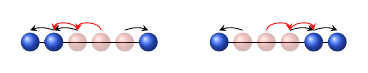}
\caption{Two distinct external paths leading to the same shortest-distance configurations for fermions with $q_{B}=3$ when $L = 6$. \label{L6}} 
\end{figure*}

We can utilize the physical picture of ECW to derive several spectral properties of the correlation matrix. For free fermions, it is well known that entanglement-related quantities can be efficiently computed by diagonalizing the correlation matrix of the subsystem. This powerful simplification arises because the many-body wavefunction of noninteracting fermions is a Slater determinant, which always satisfies Wick’s theorem \cite{peschel2004,peschel2003}. As a result, the reduced density matrix of subsystem $A$ takes a Gaussian form:
\bea
\rho_{A}=\frac{\mathrm{e}^{-c_{m}^{\dagger}h_{mn}c_{n}}}{\mathrm{Tr}\mathrm{e}^{-c_{m}^{\dagger}h_{mn}c_{n}}}=\frac{\mathrm{e}^{-a_{k}^{\dagger}\epsilon_{k}a_{k}}}{\mathrm{Tr}\mathrm{e}^{-a_{k}^{\dagger}\epsilon_{k}a_{k}}},\label{RDM}
\eea
where $c_{m}$ is the fermionic annihilation operator at site $m$ restricted in subsystem $A$, $h$ is single-particle entanglement Hamiltonian, $\epsilon_{k}$ is the $k$-th eigenvalue of $h$ and $a_{k}$ is the corresponding fermionic eigenmode. The matrix $h$ is determined by the equal-time correlation matrix of subsystem $A$, defined as
\bea
C_{A,mn}=\left\langle c^{\dagger}_{m}c_{n}\right\rangle.\label{CM}
\eea
From Eqs.~\eqref{RDM} and \eqref{CM}, their relationship can be derived as
\bea
h^{T}=\ln(I-C_{A})/C_{A},
\eea
which implies the spectrum of the correlation matrix $C_{A}$ (denoted as $\lambda_{k}^{C}$) is related to the single-particle entanglement spectrum $\epsilon_{k}$ via:
\bea
\lambda^{C}_{k}=\frac{1}{1+\mathrm{e}^{\epsilon_{k}}}.\label{lamC}
\eea
The spectrum of reduced density matrix $\rho_{A}$, denoted as $\lambda^{\rho}$, can then be written as a function of the occupation numbers $n_{k}=a^{\dagger}_{k}a_{k}$:
\bea
\lambda^{\rho,\left\{n_{1},n_{2},...\right\}}=\prod_{k}(\lambda^{C}_{k})^{n_{k}}(1-\lambda^{C}_{k})^{1-n_{k}}.\label{lamrho}
\eea

Now consider the noninteracting Hamiltonian in Eq.~\eqref{dFH} with PBCs, which exhibits the ECW phenomenon and can be understood through the shortest-distance configuration picture. At $t=0$, the correlation matrix is the identity matrix and all the eigenvalues are simply given by 1. As $\lambda^{C}_{k}$ does not exceed 1 by \eqref{lamC}, the leading behavior of $\lambda^{C}_{k}$ for small $t$ must take the form
\bea
\lambda^{C}_{k}(t)=1-f_{k}(Jt)^{l_{k}}+\mathcal{O}((Jt)^{l_{k}+1}),\label{LamC}
\eea
where $f_{k}\ge0$, and the exponents $l_{k}$ are assumed to increase monotonically with $k$. As previously discussed, the dominant eigenvalue of $\rho_{A}$ is determined by the parity of $q_{B}=L/2-\sum_{k}n_{k}\neq L/2$. When $q_{B}$ is odd, there are two leading eigenvalues that are degenerate at lowest order of $t$:
\bea
\lambda^{\rho,q_{B}}_{1,2}=g_{q_{B}}(Jt)^{2d_{q_{B}}}+\mathcal{O}(t^{2d_{q_{B}}+1}),
\eea
where $d_{q_B}$ is the shortest hopping distance defined in the main text, and $g_{q_B}>0$ is a coefficient determined by the Taylor expansion of $\mathrm{e}^{-\mathrm{i}Ht}$ and the corresponding fermionic tunneling paths. In contrast, for even $q_{B}$, there is only one such dominant eigenvalue. The distance $d_{q_{B}}$ is classified by the parity of $q_{B}$:
\bea
d_{q_{B}}= \begin{cases} 
\left( \frac{q_B - 1}{2} \right)^2 + \left( \frac{q_B + 1}{2} \right)^2, & q_B \text{ is Odd} \\ 
\left( \frac{q_B}{2} \right)^2 \times 2, & q_B \text{ is  Even}
\end{cases}.
\eea

One way to understand the distance formula is as follows: when $q_B$ is odd, the particle cluster on the left (right), consisting of $(q_B - 1)/2$ particles, shifts $(q_B - 1)/2$ sites to the left (right); simultaneously, the cluster on the right (left), containing $(q_B + 1)/2$ particles, moves $(q_B + 1)/2$ sites to the right (left). When $q_B$ is even, the cluster on the left (right), composed of $q_B/2$ particles, shifts $q_B/2$ sites to the left (right).

To compute the prefactor $g_{q_B}$, we distinguish between two types of particle movement paths contributing to the configuration:
\begin{itemize}
\item[(i)] External paths, which describe the collective motion of each particle cluster, treated as a composite object, as it shifts across the lattice;
\item[(ii)] Internal paths, which account for the relative motion of particles within a given cluster.
\end{itemize}

In the generic case, the external path to the same shortest-distance configuration is unique, and only the number of internal paths needs to be considered. The only exception arises when $q_{B}=L/2$ is odd, in which case there are two distinct external paths leading to the same configuration (see Fig.~\ref{L6}). Accordingly, the number of external paths $\chi^{E}(q_{B})$ is given by
\bea
\chi^{E}(q_{B}) = \begin{cases} 
2, & q_{B}=L/2 \text{ is Odd} \\ 
1, & \text{Otherwise}.
\end{cases}
\eea

We define $\chi^{I}(N)$ as the number of internal paths within a cluster of $N$ particles. Computing $\chi^{I}(N)$ is equivalent to analyzing a combinatorial process: initially, $N$ particles occupy sites $1, 2, \dots, N$ on a one-dimensional lattice. Each particle must move exactly $N$ steps to the right such that their final positions are $N+1, N+2, \dots, 2N$, respectively. At each time step, only one particle is allowed to move one site to the right. During the entire process, particles are prohibited from occupying the same site or moving backward. This setup is mathematically equivalent to enumerating the number of standard Young tableaux of shape $N\times N$ (see Fig.~\ref{youngtableaux}). The number of such tableaux, and hence the number of valid internal paths, is given by the hook-length formula \cite{ma2007}:
\bea
\chi^{I}(N)=\frac{(N^2)!}{\prod _{i=0}^{N-1}(N+i)!/i!}.
\eea

\begin{figure}[h]
    \centering
    \[
    \begin{array}{|c|c|c|}
    \hline
    1 & 2 & 5 \\
    \hline
    3 & 4 & 7 \\
    \hline
    6 & 8 & 9 \\
    \hline
    \end{array}
    \]
    \caption{An example of a $3\times3$ standard Young tableau. Particles are restricted to rightward motion. The $i$-th row corresponds to the $i$-th particle (from right to left), and the $j$-th column indicates the $j$-th rightward step taken. The entry in each cell labels the order in which steps occur globally.}
    \label{youngtableaux}
\end{figure}

\begin{figure*}[!t]
\includegraphics[width=\textwidth, height=\textheight, keepaspectratio]{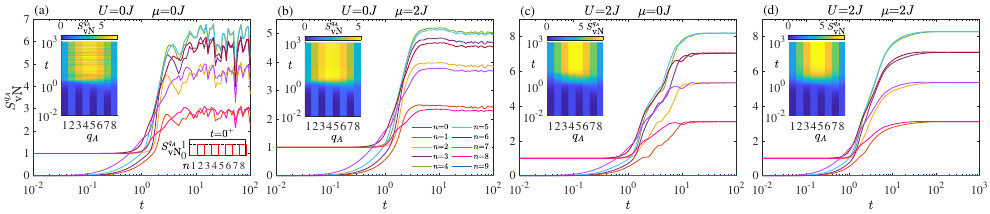}
\caption{(a)-(d) Time evolution of the symmetry-resolved von Neumann entropy $S_{\text{vN}}^{q_A}$ (in units of $\log 2$) for the bosonic half-chain entanglement entropy at system size $L = 18$. The system is initialized in a domain-wall configuration, and the entropy is resolved across symmetry sectors $q_A = 0$ to $9$. Each panel corresponds to a different choice of interaction strength $U$ and disorder strength $\mu$. A characteristic ECW pattern is observed, which breaks down at different times across channels: sectors with $S_{\text{vN}}^{q_A} = \log 2$ remain intact up to times of order $\mathcal{O}(10^0)$, while those with $S_{\text{vN}}^{q_A} = 0$ begin to deviate as early as $\mathcal{O}(10^{-1})$. For the disordered case, results are averaged over 200 independent disorder realizations.\label{bose}} 
\end{figure*}

\begin{figure*}[!t]
\includegraphics[width=0.8\textwidth, keepaspectratio]{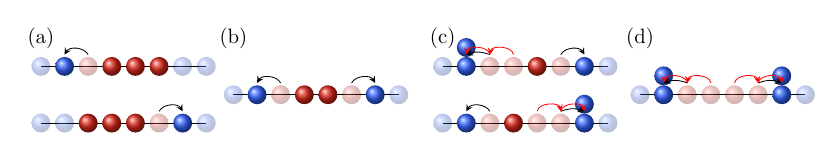}
\caption{(a)-(d) Shortest-distance configurations for bosons with total system size $L = 8$ and subsystem symmetry charges $q_A = 3, 2, 1, 0$, respectively. Red (blue) balls represent particles located in subsystem $A$ ($B$). In contrast to fermions, bosons are not subject to the Pauli exclusion principle. As a result, in their shortest-distance configurations, bosons tend to cluster massively at the outermost sites of subsystem $B$. \label{L8_Bose}}
\end{figure*}

As previously discussed, the leading contribution in the Taylor expansion of the time-evolution operator $\mathrm{e}^{-\mathrm{i}Ht}$ arises from the term $(-\mathrm{i}Ht)^{d_{q_{B}}}/d_{q_{B}}!$. Among the total of $d_{q_B}$ Hamiltonians in this expansion, \( d_{q_B}^L \) are responsible for the leftward propagation of the left cluster, and \( d_{q_B}^R \) govern the rightward propagation of the right cluster. 
\begin{table}[h]
    \centering
    \caption{Theoretical and approximate values of \( l_k \) and \( f_k \) for \( L=14 \). For \( k=7 \), the approximate value of \( f_k \) underestimates the theoretical result by a factor of 2.}
    \label{lkfk}
    \begin{tabular}{cccccccc}
        \toprule
        \( k \) & 1 & 2 & 3 & 4 & 5 & 6 & 7 \\
        \midrule
        Theo. \( l_k \) & 2 & 2 & 6 & 6 & 10 & 10 & 14 \\
        Approx. \( l_k \) & 2 & 2 & 6 & 6 & 10 & 10 & 14 \\
        Theo. \( f_k \) & 1 & 1 & \( \frac{1}{144} \) & \( \frac{1}{144} \) & \( \frac{1}{518400} \) & \( \frac{1}{518400} \) & \( \frac{1}{2540160000} \) \\
        Approx. \( f_k \) & 1 & 1 & \( \frac{1}{144} \) & \( \frac{1}{144} \) & \( \frac{1}{518400} \) & \( \frac{1}{518400} \) & \( \frac{1}{5080320000} \) \\
        \bottomrule
    \end{tabular}
\end{table}

Combining these contributions with the Taylor coefficient, the total prefactor becomes $1/(d_{q_B}^L!d_{q_B}^R!)$. Taking all relevant factors into account, the coefficient $g_{q_{B}}$ takes the following form:
\beq
g_{q_{B}}=\begin{cases} 
(\frac{\chi^{I}(q_{B}/2)\chi^{I}(q_{B}/2)}{(q_{B}^2/4)!(q_{B}^2/4)!})^2, & q_{B}\text{ is Even} \\ 
(2\frac{\chi^{I}((q_{B}-1)/2)\chi^{I}((q_{B}+1)/2)}{((q_{B}-1)^2/4)!((q_{B}+1)^2/4)!})^2, & q_{B}=L/2 \text{ is Odd} \\
(\frac{\chi^{I}((q_{B}-1)/2)\chi^{I}((q_{B}+1)/2)}{((q_{B}-1)^2/4)!((q_{B}+1)^2/4)!})^2, & \text{Otherwise}
\end{cases}.
\eeq

Here, the squaring originates from the fact that diagonal elements of the reduced density matrix correspond to squared amplitudes of the wavefunction.

From Eq.~\eqref{lamrho}, we see that the additional power of $t$ in $\lambda^{\rho,Q_B+1}$ relative to $\lambda^{\rho,Q_B}$ comes from the factor $1 - \lambda^{C}_{Q_B+1}$. Consequently, the scaling with $t$ and the corresponding coefficient from Eq.~\eqref{LamC} are given by
\bea
l_{k}=4\left[ \frac{k-1}{2} \right]+2\label{dQB}
\eea
and
\bea
f_{k}=\begin{cases}
   4 \frac{([(k-1)/2]!)^4}{\left\{(2[(k-1)/2])!(2[(k-1)/2]+1)!\right\}^2}, & k=L/2 \text{ is Odd} \\
   \frac{([(k-1)/2]!)^4}{\left\{(2[(k-1)/2])!(2[(k-1)/2]+1)!\right\}^2}, & \mathrm{Otherwise} \\ 
\end{cases}. \label{lQB}
\eea
where $\left\lfloor x \right\rfloor$ denotes the floor function (i.e., the greatest integer less than or equal to $x$). Importantly, when $k < L/2$ is odd, one finds that $l_k = l_{k+1}$ and $f_k = f_{k+1}$, indicating a pairwise degeneracy in the lowest nontrivial order (beyond the constant term) of the eigenvalues in the $t$-expansion of the correlation matrix.
 
To verify the above analysis, we consider the clean tight-binding model given by Eq.~\eqref{dFH} with $U = \mu_i = 0$. The time evolution of the annihilation operator can be computed via the Jacobi–Anger expansion: 
 \beq
 \begin{aligned}
     c_{m}(t)&=\frac{1}{\sqrt{L}} \sum_{k}\mathrm{e}^{\mathrm{i}km-2\mathrm{i}t\cos k}c_{k}\\
&=\frac{1}{L} \sum_{j}\sum_{k}\mathrm{e}^{\mathrm{i}k(m-j)-2\mathrm{i}t\cos k}c_{j}\\
&=\frac{1}{L} \sum_{j}\sum_{k}\sum_{l=-\infty}^{\infty}(-\mathrm{i})^{l}J_{l}(2Jt)\mathrm{e}^{\mathrm{i}k(m-j)}\mathrm{e}^{\mathrm{i}kl}c_{j}\\
&=\sum_{j}\sum_{M\in \mathbb{Z}}(-\mathrm{i})^{j-m+ML}J_{j-m+ML}(2Jt)c_{j}.
 \end{aligned}\label{anop}
 \eeq
From this, the correlation matrix element is computed as
\beq
\begin{aligned}
    \left\langle c^{\dagger}_{n}(t)c_{m}(t)\right\rangle
    &=\sum_{j=1}^{L/2}\sum_{N\in \mathbb{Z}}\sum_{M\in \mathbb{Z}}(-\mathrm{i})^{n-NL-m+ML}\\
    &\times J_{j-n+NL}(2Jt)J_{j-m+ML}(2Jt).
\end{aligned}\label{CM2}
\eeq

We have verified that the eigenvalue spectrum of this correlation matrix, at leading order in $t$, is consistent with the scaling forms given in Eqs.~\eqref{lamC}, \eqref{dQB}, and \eqref{lQB}. Keeping only the $N = 0$, $M = 0$ terms in Eq.~\eqref{CM2} corresponds to the continuum approximation $\sum_k/L \to \int \mathrm{d}k / (2\pi)$ in Eq.~\eqref{anop}. Within this approximation, symbolic computation tools such as Mathematica can be used to solve the secular equation and extract the leading-order scaling behavior of the eigenvalues in $t$.

In most cases, this continuum approximation accurately reproduces both the scaling exponent and the coefficient of $t$. However, when $k = L/2$ is odd, the approximate coefficient is smaller by a factor of 2 compared to the theoretical value. This discrepancy is resolved by including the $N = \pm 1$, $M = \pm 1$ terms in Eq.~\eqref{CM2}.

Although we have explicitly verified these results in the clean system, we emphasize that the scaling behavior described by Eqs.~\eqref{lamC}, \eqref{dQB}, and \eqref{lQB} remains robust in the presence of disorder in the chemical potential.

\section{Universality of the entanglement channel wave and its SU(2) generalization} \label{sec3}

In this section, we demonstrate the universality of the ECW phenomenon across different classes of quantum many-body systems. Beyond spinless fermions, the ECW pattern also emerges in bosonic systems and spinful fermionic systems that respect SU(2) symmetry. 

\subsection{The entanglement channel wave in bosonic systems}

To explore the ECW phenomenon in bosonic systems, we consider the one-dimensional disordered Bose-Hubbard model \cite{gurarie2009} with PBCs:
\begin{align}
\hat{H}_{\text{dBH}}=&-J\sum_{i}(b_{i}^{\dagger}b_{i+1}+\text{H.c.})+\sum_{i}\mu_{i}n_{i}\nonumber\\
&+\sum_{i}\frac{U}{2}n_{i}(n_{i}-1),\label{dBH}
\end{align}
where $b^{\dagger}_{i}$ creates a boson at site $i$, $n_{i}=b^{\dagger}_{i}b_{i}$ is the number operator, $\mu_{i}$ is the site-dependent disordered chemical potential, and $U$ is the on-site interaction strength. 
 
Analogous to the fermionic case, we study the time evolution of SREE following a quantum quench from a domain-wall initial state. We explore four representative parameter regimes, including both clean and disordered, interacting and noninteracting settings. For disordered systems, we perform statistical averaging over 200 independent disorder realizations to obtain reliable SREE profiles within each particle number sector.

\begin{figure*}[!t]
\includegraphics[width=\textwidth, height=\textheight, keepaspectratio]{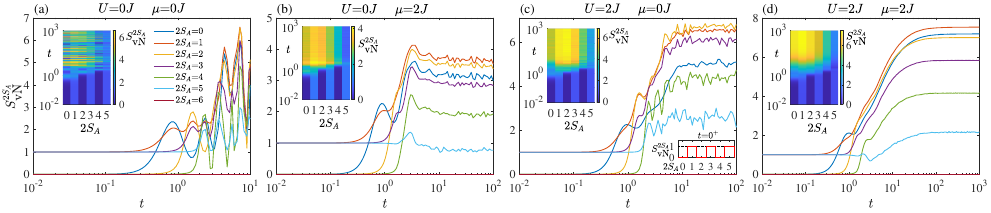}
\caption{(a)-(d) Time evolution of the SU(2)-resolved von Neumann entanglement entropy $S_{\text{vN}}^{S_{A}}$ (in units of $\log 2$) for the half-chain ($L=12$) in the spinful Fermi-Hubbard model, computed from a generalized domain-wall initial state across symmetry sectors $2S_{A} = 0$ to $6$. Each panel corresponds to a different set of interaction strength $U$ and disorder amplitude $\mu$. The ECW patterns break sequentially, starting from lower to higher spin sectors, with breakdown occurring between $\mathcal{O}(10^{-1})$ and $\mathcal{O}(10^{0})$ timescales.\label{spinfulfermion}} 
\end{figure*}

In clean, noninteracting bosonic systems, the reduced density matrix exhibits a spatial reflection symmetry similar to Eq.~\eqref{reflection}, though lacking the fermionic sign structure due to the absence of anticommutation relations. Despite the distinct quantum statistics, the shortest-distance configurations in bosonic systems display structural similarities to their fermionic counterparts in terms of combinatorics and pairings (see Fig.~\ref{L8_Bose}). As a result, the ECW pattern emerges naturally at short times, even in the bosonic setting.

A key distinction arises from the absence of the Pauli exclusion principle: bosons are allowed to occupy the same site. In their shortest-distance configurations, all bosons can cluster at the outermost sites of subsystem $B$, resulting in an ECW pattern that, unlike in the fermionic case, is insensitive to the size of subsystem $B$. This illustrates how the ECW can manifest in systems with fundamentally different microscopic rules, reinforcing its universality across different particle statistics.

\subsection{SU(2) generalization of the ECW pattern}

\begin{figure*}[!t]
\includegraphics[width=0.8\textwidth, keepaspectratio]{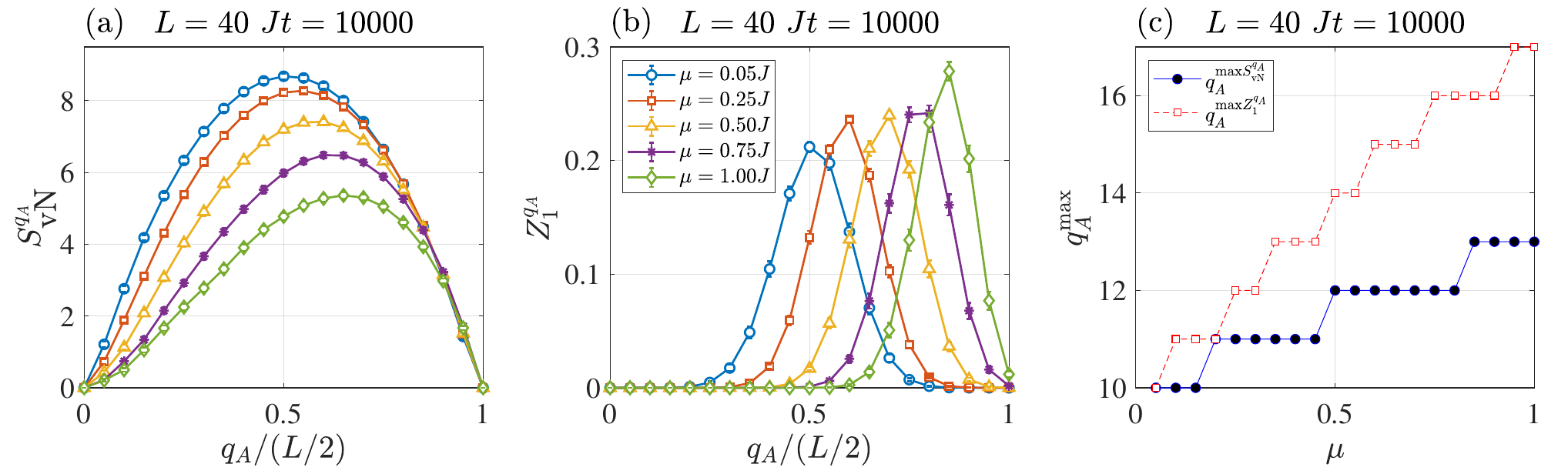}
\caption{(a) Symmetry-resolved von Neumann entropy $S_{\mathrm{vN}}^{q_A}$ (in units of $\log 2$) for each charge sector $q_A$, plotted for several disorder strengths $\mu$. (b) Probability $Z_1^{q_A}$ of the many-body wavefunction projected onto each sector $q_A$, also shown for various $\mu$. (c) Charge sectors corresponding to the maximal values of $S_{\mathrm{vN}}^{q_A}$ and $Z_1^{q_A}$, denoted by $q_A^{\mathrm{max} S_{\mathrm{vN}}^{q_A}}$ and $q_A^{\mathrm{max} Z_1^{q_A}}$, respectively, as functions of disorder strength $\mu$. All results are for system size $L=40$ and time $Jt = 10000$, averaged over 100 disorder realizations. \label{L40}} 
\end{figure*}

In this section, we extend our study of ECW to spinful fermionic systems with full SU(2) spin-rotational symmetry. Specifically, we investigate the one-dimensional disordered Fermi-Hubbard model with PBCs, defined on a lattice of even length $L$. The Hamiltonian reads:
\begin{align}
\hat{H}_{\text{sFH}}&=J\sum_{i,s=\uparrow,\downarrow}( c_{is}^\dagger c_{i+1s}+\text{H.c.}) -\sum_{i}\mu_{i}(n_{i\uparrow}+n_{i\downarrow})\nonumber \\
&+\sum_{i}U_{i} n_{i\uparrow} n_{i\downarrow},\label{sdFH}
\end{align}
where $c^\dagger_{is}$ ($c_{is}$) creates (annihilates) a fermion of spin $s$ on site $i$, and $n_{is} = c^\dagger_{is} c_{is}$ is the local number operator. The model possesses an SU(2) spin-rotational symmetry generated by $S^\alpha = \sum_i c^\dagger_{is} (\sigma^\alpha_{ss'}/2) c_{is'}$ for $\alpha = x, y, z$, where $\sigma^\alpha$ are the Pauli matrices \cite{essler2005}. We consider a generalized domain-wall initial state:
$\left|\psi(0)\right\rangle=\left|\bigotimes_{i=1}^{L/2}(\uparrow\downarrow)_{i}\bigotimes_{i=L/2+1}^{L}0_{i}\right\rangle$, in which the left half of the chain ($A$) is fully occupied with spin-up and spin-down fermions, while the right half ($B$) is empty. This state is a global spin singlet, satisfying $S = 0$ and $S^z = 0$, and the time-evolved density matrix $\rho(t) = |\psi(t)\rangle\langle\psi(t)|$ remains SU(2)-invariant, i.e., $[S^\alpha, \rho(t)] = 0$. Owing to this symmetry, the Hilbert space of the full system decomposes into SU(2)-invariant subspaces. For the bipartition into subsystems $A$ (sites $1$ to $L/2$) and $B$ (sites $L/2+1$ to $L$), the relevant decomposition is: $\mathcal{H}^{S=0,S^{z}=0}=\bigoplus_{S_{A}=0}^{L/4} \bigoplus_{S_{A}^{z}=-S_{A}}^{S_{A}}\mathcal{H}_{A}^{S_{A},S_{A}^{z}}\bigotimes\mathcal{H}_{B}^{S_{A},-S_{A}^{z}}$, where subsystems $A$ (sites $1$ to $L/2$) and $B$ (sites $L/2+1$ to $L$) are defined. This follows from the fact that the singlet representation appears in the tensor product of any SU(2) representation with its conjugate (which, for SU(2), is itself). For comparison, if only U(1) symmetry (i.e., conservation of $S^z$) is preserved, the decomposition becomes: $\mathcal{H}^{S^{z}=0}=\bigoplus_{S_{A}^{z}=-L/4}^{L/4}\mathcal{H}_{A}^{S_{A}^{z}}\bigotimes\mathcal{H}_{B}^{-S_{A}^{z}}$. Tracing out subsystem $B$, the reduced density matrix $\rho_{A}(t)$ commutes with both $S_A^\alpha$ and $S_A^2$, leading to the block-diagonal form: $\rho_{A}(t)=\bigoplus_{S_{A},S_{A}^{z}}\rho_{A,S_{A},S_{A}^{z}}(t)$. Under U(1) symmetry alone, it is block-diagonal in $S_{A}^{z}$: $\rho_{A}(t)=\bigoplus_{S_{A}^{z}}\rho_{A,S_{A}^{z}}(t)$. Under SU(2) symmetry, Schur’s lemma implies that $\rho_{A, S_A, S_A^z}(t)$ depends only on $S_A$, and not on $S_A^z$. The normalized reduced density matrix is $\tilde{\rho}_{A,S_A, S_A^z}=\rho_{A,S_A, S_A^z}/\mathrm{Tr}\rho_{A,S_A, S_A^z}$. The symmetry-resolved Rényi and von Neumann entropies for singlet states are then given by \cite{goldstein2018}: $S^{S_{A},S_{A}^{z}}_{\alpha}(t)=\ln\mathrm{Tr}(\tilde{\rho}^{\alpha}_{A,S_A, S_A^z}(t))/(1-\alpha) $ and $S^{S_{A},S_{A}^{z}}_{\text{vN}}(t)=-\mathrm{Tr}\tilde{\rho}^{\alpha}_{A,S_A, S_A^z}(t)\log\tilde{\rho}^{\alpha}_{A,S_A, S_A^z}(t)$. (For states with $S\neq0$, the definition of the SU(2) symmetry-resolved entanglement entropy can be found in \cite{bianchi2024}.) A practical method to extract SU(2)-resolved quantities from U(1) ones is provided in Refs.~\cite{goldstein2018,calabrese2021}, based on the identity: $\mathrm{Tr} \rho_{A,S_{A},S_{A}^{z}}^n = \mathrm{Tr} \rho_{A,S_{A}^{z}=S_{A}}^n - \mathrm{Tr} \rho_{A,S_{A}^{z}=S_{A}+1}^n
$. For $S_{A}^{z}=S_{A}\geq0$, the leading contribution to the reduced density matrix block $\rho_{A,S_{A}^{z}}$ at short times $t$ arises from configurations in which $2S_{A}^{z}$ spin-down fermions tunnel from subsystem $A$ to $B$. Among these, the shortest-distance configurations---those requiring the minimal number of hopping events---dominate the dynamics in the early-time limit. These configurations are structurally similar to those found in the spinless fermion case, except that here only spin-down fermions are involved in the tunneling process. Importantly, fewer particles need to tunnel out of subsystem $A$ in the $S_{A}^{z}=S_{A}$ channel compared to the $S_{A}^{z}=S_{A}+1$ channel. As a result, the normalized trace of the reduced density matrix in the SU(2) channel can be well approximated by $\mathrm{Tr} \tilde\rho_{A,S_{A},S_{A}^{z}}^n\approx \mathrm{Tr} \tilde\rho_{A,S_{A}^{z}=S_{A}}^n
$, where $\tilde\rho_{A,S_{A}^{z}=S_{A}}=\rho_{A,S_{A}^{z}=S_{A}}/\mathrm{Tr}\rho_{A,S_A^z=S_{A}}$. For $S_{A}\neq L/4$ (recall that the spin-$L/4$ representation in $\mathcal{H}_{A}$ is unique and thus contributes trivially), the shortest-distance configurations exhibit two key features: 
\begin{enumerate}
    \item Multiplicity: The number of such configurations in the \( S_A^{z}=S_{A} \) sector is \( 2^{(1-(-1)^{2S_{A}})/2} \).
    \item Pairing: The configurations \( \{\alpha^s \in \mathcal{H}_A^{S_{A}^{z}=S_{A}}, \beta^s \in \mathcal{H}_B^{S_{B}^{z}=-S_{A}}\} \) form a one-to-one correspondence between $\alpha^s$ and $\beta^s$. 
\end{enumerate}
These properties determine the structure of  $\tilde\rho_{A,S_{A}^{z}=S_{A}}$ in the $t\rightarrow0^{+}$ limit:
\begin{itemize}
    \item For integer \(S_{A} \), \( \tilde\rho_{A,S_{A}^{z}=S_{A}}(0^+) \) possesses a single diagonal element equal to 1.
    \item For half-integer \( S_{A} \), \( \tilde\rho_{A,S_{A}^{z}=S_{A}}(0^+) \) it contains two diagonal elements, each equal to \( 1/2 \).
\end{itemize}

Consequently, the SU(2)-resolved Rényi and von Neumann entropies at $t=0^{+}$ (for any Rényi index $\alpha$ and $S_{A}\neq L/4$) are given by:
\bea
S^{S_{A},S_{A}^{z}}_{\alpha}(0^{+})=\frac{1-(-1)^{2S_{A}}}{2}\log 2.
\eea
This expression reveals a parity-induced quantization of entanglement: the entropy is $\log2$ for half-integer $S_{A}$ ,indicating two equally probable basis states, and zero for integer $S_{A}$, indicating a pure-state contribution. This quantized behavior parallels that observed in the U(1)-symmetric case. 

We numerically confirm that the ECW pattern also emerges in the SU(2)-symmetric system at short times (see Fig.~\ref{spinfulfermion}). For spinful free fermions (i.e., Eq.~\eqref{sdFH} with $U_{i}=0$, the correlation matrix takes the form $C_{A}=\mathrm{diag}(C_{A}^{\uparrow\uparrow},C_{A}^{\downarrow\downarrow})$, where both $C_{A}^{\uparrow\uparrow}$ and $C_{A}^{\downarrow\downarrow}$ follow the same structure as in the spinless case described by Eqs.~\eqref{lamC}, \eqref{dQB}, and \eqref{lQB}.

\subsection{Long-time limit of the entanglement channel wave}

For spinless fermions, the ECW pattern exhibits a gradual temporal degradation across all symmetry channels within a characteristic timescale of $\mathcal{O}(10^0)$ to $\mathcal{O}(10^1)$. Channels with higher particle numbers tend to lose the ECW structure earlier than those with lower particle occupancy (see Fig.~\ref{fermi}). In the clean, noninteracting case, as shown in the insets at the upper-left corners of each panel in Fig.~\ref{fermi}, the SREE does not approach a steady-state value at long times. In contrast, the disordered spinless fermion system exhibits non-Gaussian behavior, most clearly reflected in the asymmetry of $S_{\mathrm{vN}}^{q_A}(t) \neq S_{\mathrm{vN}}^{L/2 - q_A}(t)$.
This asymmetry is absent in the interacting case. Previous work has suggested that the appearance or suppression of such non-Gaussian features can serve as a diagnostic of thermalization \cite{chen2023c}. 

To explore this finding further in the disordered free-fermion regime, we employ the correlation matrix method on a larger system to get rid of the finite-size effect. However, evaluating SREE in small symmetry sectors $q_A$ becomes increasingly challenging due to the extremely low values of the corresponding projected weight, defined as $Z_1^{q_A}(t) \equiv \mathrm{Tr} \rho_{A,q_A}(t)$, which represents the probability of finding subsystem $A$ in the sector with particle number $q_{A}$. In strongly disordered systems, $Z_1^{q_A}(t)$ can become vanishingly small. This is attributed to disorder-induced Anderson localization \cite{anderson1958,thouless1972a}, which reduces the mean free path and causes most particles to remain confined within subsystem $A$ [see Fig.~\ref{L40}(b)]. Given this limitation, we focus on a system of size $L=40$ and systematically vary the disorder strength $\mu$ (see Fig.~\ref{L40}). At weak disorder, both the SREE and sector probabilities exhibit near-symmetric behavior, indicating a near-complete loss of memory of the initial state. Specifically, we observe $S_{\mathrm{vN},A}^{q_A} \approx S_{\mathrm{vN},A}^{L/2 - q_A} = S_{\mathrm{vN},B}^{q_B}$ and $Z_{1,A}^{q_A} \approx Z_{1,A}^{L/2 - q_A} = Z_{1,B}^{q_B}$. As disorder increases, both the entanglement-dominant sector $q_A^{\mathrm{max} S_{\mathrm{vN}}^{q_A}}$ and the probability-dominant sector $q_A^{\mathrm{max} Z_1^{q_A}}$ shift away from the center toward larger $q_A$. Notably, the peak entanglement decreases with increasing disorder, while the peak probability increases.

Interestingly, we find a notable entanglement-probability separation: as disorder strengthens, the shift of the entanglement-maximizing sector toward larger $q_{A}$ occurs more rapidly than that of the probability-maximizing sector [see Fig.~\ref{L40}(c)]. This decoupling highlights a nontrivial distinction between where entanglement is concentrated and where probability is most likely to be found in the presence of strong disorder.

As shown in Fig.~\ref{bose}, the ECW pattern in the bosonic system persists robustly up to timescales on the order of $\mathcal{O}(10^{-1})$, confirming the early-time stability of parity-based quantization in the dynamics. The breakdown of the ECW pattern occurs at different times depending on the parity of the charge sector: channels with $S_{\mathrm{vN}}^{n} = \log 2$ retain the ECW structure up to $\mathcal{O}(10^{0})$, whereas those with $S_{\mathrm{vN}}^{n} = 0$ begin to degrade around $\mathcal{O}(10^{-1})$. In contrast to the fermionic case, the SREE in bosonic systems can be stabilized purely by interactions, even in the absence of disorder. Furthermore, no significant non-Gaussian features are observed, even when disorder is present. While disorder introduces slight deviations from Gaussian behavior at long times, these effects remain much weaker than those found in spinless fermionic systems.

In the SU(2)-symmetric case, our results show that ECW patterns corresponding to small spin sectors consistently degrade earlier, irrespective of the presence of disorder or interactions. In the noninteracting disordered regime, the SU(2) SREE exhibits a distinctive overshooting phenomenon before reaching equilibrium---a feature not observed in the U(1) or bosonic settings. Furthermore, in the long-time limit, the SU(2) SREE exhibits a strongly non-Gaussian distribution, with small spin sectors developing substantially stronger entanglement than larger ones.

\section{Conclusion}\label{sec4}

In summary, we have investigated the quench dynamics of U(1)- and SU(2)-symmetry-resolved entanglement entropy using both the Krylov-subspace iterative method and the correlation matrix technique. In the short-time regime, we identified a universal phenomenon---the entanglement channel wave---which is remarkably robust against variations in interaction strength, disorder strength, and quantum statistics. For free fermions, the ECW structure further enabled us to derive analytical properties of the correlation matrix spectrum.

The melting sequence of the ECW pattern varies across different systems:
\begin{itemize}
    \item In fermionic systems, U(1) ECW channels with higher particle numbers degrade earlier.
    \item In bosonic systems, ECW breakdown proceeds sequentially from the $S_{\text{vN}}^{q_{A}}=0$ channels to those with $S_{\text{vN}}^{q_{A}}=\log2$.
    \item In SU(2)-symmetric systems, ECW patterns in smaller spin sectors tend to deteriorate before those in larger ones.
\end{itemize}

At long times, the SREE of clean free fermions typically exhibits chaotic dynamics. In U(1)-symmetric systems, the SREE tends to show Gaussian behavior, except in one striking case: for disordered free spinless fermions, the SREE demonstrates an entanglement-probability separation, where the entanglement-maximizing sector shifts more rapidly than the probability-maximizing sector. In the SU(2)-symmetric case, small spin sectors often retain higher entanglement than larger ones even at late times.

Understanding and harnessing the microscopic origins of these symmetry-resolved dynamical features---especially the mechanisms behind ECW formation, breakdown, and long-time entanglement asymmetry---remains an open and compelling direction for future research.

\bibliography{ECW}

\end{document}